\documentclass[epj]{svjour}
\usepackage{psfig}
\begin{document}

\title{Production of $a_0$-mesons in the reactions $\pi N \to a_0 N$
and $p p \to d a_0^+$ at GeV energies \thanks{Supported by DFG, RFFI
and GSI Darmstadt}}
\author{V. Yu. Grishina $^a$, L. A. Kondratyuk $^b$, E. L. Bratkovskaya $^c$,
M. B\"uscher $^d$, and W. Cassing $^c$ \\}
\institute{$^a$ Institute for Nuclear Research,
     60th October Anniversary Prospect 7A,  117312 Moscow, Russia \\
$^b$ Institute of Theoretical and Experimental Physics,
     B.Cheremushkinskaya 25, 117259 Moscow, Russia \\
$^c$ Institut f\"ur Theoretische Physik, Universit\"at Giessen,
     D-35392 Giessen, Germany\\
$^d$ Institut f\"ur Kernphysik, Forschungszentrum J\"ulich,
     D-52425 J\"ulich, Germany}
\date{Received: date / Revised version: date}

\abstract{
We investigate the reactions $\pi N \to a_0 N$ and $p p \to d
a_0^+$ near threshold and at medium energies. An effective
Lagragian approach and the Regge pole model are applied to analyze
different contributions to the cross section of the reaction $\pi
N \to a_0 N$. These results are used to calculate the differential
and total cross sections of the reaction $p p \to d a_0^+$ within
the framework of the two-step model in which two nucleons produce
an $a_0$-meson via $\pi$ -meson exchange and fuse to a deuteron.
The necessity of new measurements on $a_0$ production and
branching fractions (of its decay to the $K\bar K$ and $\pi\eta$
channels) is emphasized for clarifying the $a_0$ structure.
Detailed predictions for the reaction $pp \rightarrow d a_0^+$ are
presented for the energy regime of the proton synchrotron COSY-J\"ulich.}


\authorrunning{V. Yu. Grishina et al.}
\maketitle

\titlerunning{Production of $a_0$-mesons in the reactions $\pi N \to a_0 N$
and $p p \to d a_0^+$ at GeV energies}

\section{Introduction}

\begin{table*}[t]
\begin{center}
\begin{tabular}{c c c c c c }
\hline
Reaction & $R$ & $M_r($GeV$)$ & $g_1($GeV$)$ & Comment & Reference
\\ \hline
$p\bar p\to \eta\pi^0\pi^0, \eta\eta\pi^0$ &
1.05$\div $2.05 & 1.013$\div $1.058 & 0.241$\div $0.287 & i) & \cite{Bugg}
\\ \hline
$p\bar p\to \eta\pi^0\pi^0, \eta\eta\pi^0$ &
1.05$\div $1.45 & 1.004$\div $1.024 & 0.229$\div $0.312 & ii) & \cite{Bugg}
\\ \hline
$p\bar p\to \eta\pi^0\pi^0, \eta\eta\pi^0$ &
1.12$\div $1.37 & 0.999$\div $1.006 & 0.211$\div $0.275 & iii) & \cite{Bugg}
\\ \hline
$p\bar p\to \eta\pi^0\pi^0$ & 1.15$\pm $0.10 &
0.999$\pm 0.00$6 & 0.218$\pm $0.020 & iv) & \cite{Amsler94}
\\ \hline
\begin{minipage} [l] {3cm}
$p\bar p\to K_L K^+\pi^-,$\\
$\phantom{p\bar p\to } K_L K^-\pi^+$
\end{minipage}
&
1.03$\pm $0.4 & 0.999$\pm 0.00$2 & 0.324$\pm $0.015 & v) & \cite{Abele}
\\ \hline
$\pi^-p\rightarrow n\eta\pi^-\pi^+, n\eta\pi^0$ & 0.91$\pm $0.10
& 1.001$\div $0.0019 & 0.122$\pm $0.008 & vi) & \cite{E852}
\\ \hline
\end{tabular}
\end{center}
\caption{\label{Tab1} Parameters of the Flatt\'e parametrization
for the $a_0(980)$. Comments: i) without any external constraint;
ii) with constraint on $|a_0(980)|^2$ at half-width from the
reaction $p\bar p\rightarrow \eta\omega\pi^0$; iii) with
constraint on $|a_0(980)|^2$ at half-width from the reaction
$p\bar p\rightarrow \eta\omega\pi^0$ and contribution from a
hypothetical $a_2^\prime(1620)$ in the fit; iv) solution B with
constraint on the $a_0$ mass from the reaction $p\bar p
\rightarrow \eta\omega\pi^0$; v) with constraint that the ratio of
integrated intensities in the $K\bar K$ and $\eta\pi$ channels is
given by Eq.~(\protect\ref{eq1}); vi)  Ref.~\protect\cite{E852}
presents the value $g_{\pi\eta}=$0.243$\pm$0.015 which is related
to $g_1$ as $g_{\pi\eta}=(2/M) g_1$. }
\end{table*}

The scalar mesons play a very important role in the physics
of hadrons since they carry the quantum numbers of the vacuum.
Nevertheless, the structure of the lightest scalar mesons
$a_0(980)$ and $f_0(980)$ is not yet understood and an important
topic of hadronic physics (see e.g.
\cite{Clo,Gen,Jan,Ani,Hadron99a,Hadron99b,Close2000} and
references therein). It has been discussed that they could be
either ``Unitarized $q\bar{q}$ states'', ``Four-quark cryptoexotic
states'', $K\bar{K}$ molecules or vacuum scalars (Gribov's
minions) (see e.g. Ref. \cite{Hadron99a}). Nowadays, theory gives
some preference to the unitarized quark model proposed by
T\"ornqvist \cite{Tornqvist} (cf. \cite{Hadron99a,Hadron99b}).
However, other options cannot be ruled out so far. Since there is
a strong mixing between the uncharged $a_0(980)$ and the
$f_0(980)$ due to a coupling to $K\bar K$ intermediate states
\cite{Jan,Kerbikov}, it is important to study independently the
uncharged and charged components of the $a_0(980)$ because the
latter ones do not mix with the $f_0(980)$ and preserve their
original quark content. It is generally expected, furthermore,
that the different $a_0(980)$ production cross sections in $\pi N$
and $NN$ reactions will provide valuable information on its
internal structure.

Until now the charged components of the $a_0(980)$ have been
studied dominantly in the $\eta\pi^+$ or $\eta\pi^-$ decay
channels \cite{PDG}. Recent experimental data from the E852
Collaboration at BNL give for the charged $a_0^+$ meson a mass of
$0.9983\pm 0.0040$ GeV/c$^2$ and a width of $0.072\pm 0.0010$
GeV/c$^2$ \cite{E852}. Note, that the mass of the $a_0$ reported
by the E852 Collaboration is significantly larger than the average
value of $0.9834\pm 0.0009$ GeV/c$^2$ presented in the last issue
of the PDG \cite{PDG}.

The branching ratios to the two main $a_0$ decay channels
($\eta\pi$ and $K\bar{K}$) are still unclear: the $\eta\pi$ mode
is quoted by the PDG \cite{PDG} as `dominant' and the $K\bar{K}$
mode as `seen'. We point out, that the data from only two
experiments \cite{Deb,Abele}, where the decay of the $a_0(980)$ to
$K\bar{K}$ was observed, have been used for the PDG analysis
\cite{PDG}. The authors of Ref.~\cite{Abele} report a ratio of
branching ratios
\begin{eqnarray}
Br(\bar pp\to a_0\pi; a_0\to K\bar K) /
Br(\bar pp\to a_0\pi; a_0\to \pi \eta) \nonumber\\
= 0.23\pm 0.05. \label{eq1}
\end{eqnarray}
However, the second branching ratio taken from Ref.
\cite{Amsler94} might have a systematic uncertainty stemming from
a strong interference of the $a_0$ signal with the broad resonance
$a_0(1450)$, which has a width of about 265 MeV. As a consequence
the $a_0(980)$ maximum in the reaction $\bar p p\rightarrow
\eta\pi^0\pi^0$ might be distorted. Moreover, the invariant-mass
resolution in Refs.~\cite{Abele,Amsler94} is only $\sim 27$
MeV/c$^2$.

In another recent study \cite{WA102} the WA102 collaboration
reported the branching ratio
\begin{equation}
\Gamma(a_0\to K\bar K) / \Gamma(a_0\to \pi \eta) = 0.166\pm
0.01\pm 0.02 , \label{eq02}
\end{equation}
which was determined from the measured branching ratio for the
$f_1(1285)$-meson,
\begin{equation}
\Gamma(f_1\to K\bar K \pi) / \Gamma(f_1\to \pi \pi \eta) =
0.166\pm 0.01\pm 0.08 , \label{eq03}
\end{equation}
produced centrally in the reaction $pp \to p_f(X_0)p_s$ at 450
GeV/c. However, the authors assumed that the $f_1(1285)$ decays
effectively by 100\%   to $a_0(980) \pi$ while the PDG quotes only
a branching  $Br(f_1(1285)\to a_0(980) \pi) =0.34 \pm 0.08$.

Therefore, it is necessary to measure the branching fractions of
the two main $a_0$ decay channels ($\eta\pi$ and $K\bar K$) under
different dynamical conditions with a higher mass resolution
($\Delta m < 10$ MeV/c$^2$) and lower background in an independent
experiment. A related experiment to detect the $a_0^+$ in both
main decay modes in the reaction $pp\to da_0^+$ will be performed
at COSY (J\"{u}lich) \cite{COSY55}. An important
dynamical feature of the latter reaction is that the production of
the $a_0^+(980)$ near threshold cannot be related to an
intermediate production of the $f_1(1285)$ (see below).

In this paper we investigate the $a_0$- production cross section in
the reactions $\pi N\to a_0N$ and $pp\to da_0^+$ near threshold
and at medium energies. In Sect. \ref{sec:br} we present a short
overview on the uncertainties of the $a_0$- decay parameters
according to present knowledge. To analyze different contributions
to the cross section of the reaction $\pi N\to a_0N$ we employ an
effective Lagragian approach as well as the Regge-pole model in
Sect. 3. The results of this analysis then are used in Sect. 4 to
calculate the differential and total cross sections of the
reaction $pp\to da_0^+$ within the framework of the two-step model
(TSM), in which two nucleons produce an $a_0$-meson via
$\pi$-meson exchange and fuse to a deuteron. The TSM has been used
before in Refs. \cite{Grishina1,Grishina2} for the analysis of
$\eta$, $\eta^\prime$, $\omega$ and $\phi$ production in the
reaction $pn\to dM$ near threshold. An important difference of our
analysis here is that the $S$-wave channel in the reaction $pp\to
da_0^+$ is forbidden due to angular momentum conservation and the Pauli
principle and that this reaction is dominated near threshold by the
$P$-wave contribution.  A summary of our work is presented in Sect.
\ref{sec:Conclusion}.

\section{Models and data on the $K\bar K$ and $\pi\eta$ decay channels
of the $a_0(980)$} \label{sec:br}

Within the framework of a coupled-channel formalism  an
appropriate parametrization of the shape of the $a_0(980)$ in each
($\eta\pi$ or $K\bar K$) channel can be taken in the form proposed
by Flatt\'e \cite{Flatte},
\begin{eqnarray}
|A_i|^2 = {\rm Const} \frac{|\Gamma_i^{}(M)|\ M_r^2}
{(M^2-M_r^2)^2+M_r^2|\Gamma_{tot}^2(M)|}
\label{matrNNf}
\end{eqnarray}
where $M_r$ is the K-matrix pole, $\Gamma_{tot}(M) = \Gamma_1(M)
+\Gamma_2(M) = g_1\rho_1+g_2\rho_2$, while $g_1$ and $g_2$ are
coupling constants to the two final states and $\rho _i$ is given
by the momenta of the final particles $q_i$ as $\rho_i=2q_i/M$.
Note that molecular or ''threshold cusp'' cases would imply a
dominance of the $|K\bar K\rangle$ component in Fock space and
therefore correspond to a relatively large ratio $R=(g_2/g_1) \gg
1$. In Table \ref{Tab1} we present the most recent results for the
$a_0(980)$ parameters $R, M_r$ and $g_1$, which show a sizeable
variation especially in the coupling $g_1$ and ratio $R$,
respectively.

In Ref.~\cite{Bugg} it has been shown that, when fitting the
$\eta\pi$ mass distribution without any additional constraints,
the parameters $M_r$, $R $ and $g_1$ cannot be fixed very well.
These parameters are strongly correlated and if one of them is
moved in steps, the value of $\chi^2$ changes rather slowly, but $M_r$,
$R$ and $g_1$ move together. Thus additional constraints are used in
most fits. In Ref.~\cite{E852} a Breit-Wigner (BW) fit of the
$a_0(980)$ shape in the $\eta\pi$ channel has been performed where
the mass and width of the $a_0^+$ were determined to be
0.9964$\pm$0.0016 and 0.062$\pm$0.006 GeV/c$^2$, respectively. The two
extractions of the $a_0$ mass and width (BW and Flatt\'e) were
found to be statistically consistent. Since in a Breit-Wigner
parametrization only two parameters enter, it is not sensitive at
all to the ratio $R$. This implies that for a reliable
determination of $R$ the measurements of both channels are
necessary. Recall that two zero's of the function
$D(M)=M^2-M_r^2+i M_r(g_1 \rho_1(M)+g_2\rho_2(M))$ define two
T-matrix poles on sheets II and III where only the position of the
 pole in sheet II defines the mass ($m_0$) and width
($\Gamma_0$) of the $a_0(980)$. Note that the pole mass $m_0$ is
usually different from the resonance mass $M_r$ in Eq. (4).
According to the PDG \cite{PDG} the average value of the $a_0(980)$
mass is $0.9834\pm 0.0009$ GeV/c$^2$ for the
$\eta\pi$ final state (without the new result of the E852
Collaboration\cite{E852} ($0.9983\pm 0.004$ GeV/c$^2$)) and
$0.9808\pm 0.0027$ GeV/c$^2$ for the $K\bar K$ final state
\cite{Abele}. The width of the $a_0(980)$ is quoted as $0.092\pm 0.008$
GeV in the $K\bar K$ final state \cite{Abele} and $0.072\pm 0.01$ GeV
in the $\eta\pi$ final state \cite{E852}.

The values of the ratio $R$ presented in Table \ref{Tab1} are not
in favor of a pure molecular or pure ''threshold cusp''
interpretation of the $a_0(980)$. This statement is also in line
with the results of Ref.~\cite{Jan}, where it was shown that
the pure ''threshold cusp'' model gives an $a_0$ width of about
200 MeV, which is much larger than the experimental value.
Nevertheless, there is still a comparatively large uncertainty in
$g_1$ and $g_2$ : the values of $g_1$ may vary from 0.12 to 0.32
GeV and $R=g_2/g_1$ from 0.9 to 2.05. A better knowledge of $g_1$
and $g_2$ will help to understand the $a_0(980)$ internal
structure or its decomposition in Fock space, respectively.

\section{The reaction $\pi N\to a_0N$}
\label{sec:pi-N}

\subsection{An effective Lagrangian Approach}

\begin{figure}[t]
\centerline{\psfig{figure=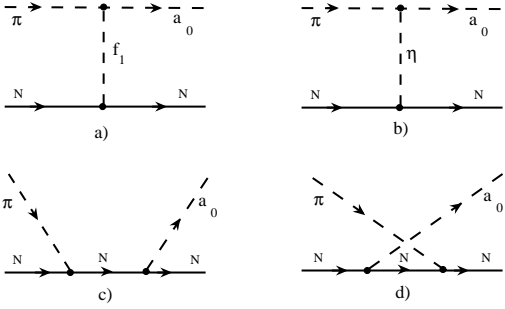,width=9cm}}
\vspace*{-1.5cm}
\caption{The diagrams for $a_0$ production in the reaction
$\pi N\rightarrow a_0 N$ near threshold. }
\label{Fig1}
\end{figure}

The most simple mechanisms for $a_0$ production in the reaction
$\pi N\rightarrow a_0 N$ near threshold  are described by the pole
diagrams shown in Fig. \ref{Fig1} a--d. It is known experimentally
that the $a_0$ couples strongly to the channels $\pi\eta$ and $\pi
f_1(1285)$ because $\pi\eta$ is the dominant decay channel of the
$a_0$ while $\pi a_0$ is one of the most important decay channels
of the $f_1(1285)$ \cite{PDG}. The amplitudes, which
correspond to the $t$-channel exchange of $\eta(550)$- and
$f_1(1285)$- mesons (a,b), can be written as
\begin{eqnarray}
&&\hspace*{-4mm}M_\eta^t(\pi^-p\rightarrow a_0^- p) = g_{\eta\pi a_0} g_{\eta NN}\
\bar u(p_2^\prime) \gamma_5 u(p_2)\nonumber\\
&\times& {1\over t-m_\eta ^2} \ F_{\eta\pi a_0}(t) F_{\eta NN}(t)
\label{eq2}\end{eqnarray}
\begin{eqnarray}
&&\hspace*{-4mm}M_{f_1}^t(\pi^- p\rightarrow a_0^- p) = g_{f_1\pi a_0} g_{f_1NN} \nonumber\\
&\times& (p_1+p_1^\prime)_\mu \ \left(g_{\mu\nu}-{q_\mu q_\nu\over
m_{f_1}^2}\right) \ \bar u(p_2^\prime) \gamma_\nu \gamma_5 u(p_2) \nonumber\\
&\times& {1\over t-m_{f_1}^2}\ F_{f_1\pi a_0}(t) F_{f_1NN}(t).
\label{eq3}
\end{eqnarray}
Here $p_1$ and $p_1^\prime$ are the four momenta of $\pi^-, a_0^-$,
whereas $p_2$ and $p_2^\prime$ are the four momenta of the initial and
final protons, respectively; furthermore, $q=p_2^\prime-p_2$,
$t=(p_2^\prime-p_2)^2$. The functions $F_j$ present form factors at the
different vertices $j$ ($j=f_1NN,\eta NN$), which are taken of the
monopole form
\begin{eqnarray}
F_j(t)=\frac{\Lambda_j^2-m_j^2}{\Lambda _j^2-t},
\label{form}
\end{eqnarray}
where $\Lambda_j$ is a cut-off parameter. In the case of $\eta$
exchange we use $g_{\eta NN}=3$, $\Lambda_{\eta NN}$=1.5 GeV from
Ref. \cite{Holinde} and $g_{\eta\pi a_0}$=2.46 GeV which results from
the width $\Gamma(a_0 \to \eta \pi$) = 80 MeV. The contribution of
the $f_1$ exchange is calculated for two parameter sets; set $A$:
$g_{f_1 NN}=11.2$, $\Lambda_{f_1 NN}=1.5$~GeV from
Ref.~\cite{Bonnf1}, set $B$: $g_{f_1 NN}=14.6$, $\Lambda_{f_1
NN}=2.0$~GeV from Ref.~\cite{Kirchbach} and $g_{f_1a_0\pi}$=2.5
for both cases. The latter value for $g_{f_1 a_0 \pi}$ corresponds
to $\Gamma(f_1\to a_0\pi)=24$~MeV and $Br(f_1\to a_0\pi)=34\%$.

\begin{figure}[t]
\psfig{figure=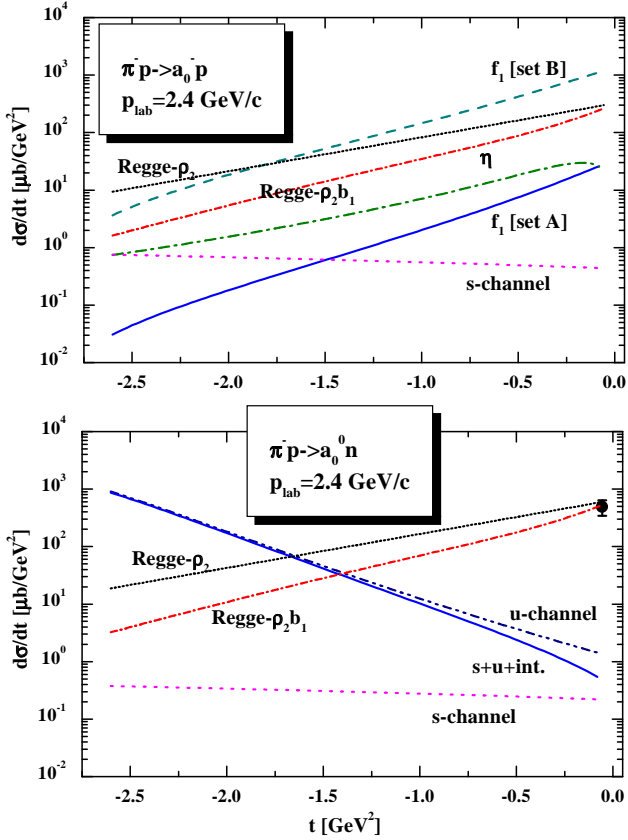,width=8.3cm}
\caption{ The differential cross sections $d\sigma/dt$ for the
reactions $\pi^-p\rightarrow  a_0^-p$ (upper part) and
$\pi^-p\rightarrow a_0^0n$ (lower part) at 2.4 GeV/c.  The
dash-dotted line corresponds to the $\eta$ exchange, solid and
dashed lines (upper part) show  the $f_1$ contributions within sets
$A$ and $B$, respectively. The dotted and dash-double-dotted lines
indicate the $s$- and $u$- channels while the solid line (lower
part) describes the coherent sum of $s$- and $u$- channel
contributions. The short dotted and short dash-dotted lines
present the results within the $\rho_2$ and ($\rho_2, \ b_1$)
Regge exchange model, respectively (see text). The experimental point
is taken from Ref.~\protect\cite{Cheshire}.}
\label{dsdt_pip}
\end{figure}

\begin{figure}[t]
\centerline{\psfig{figure=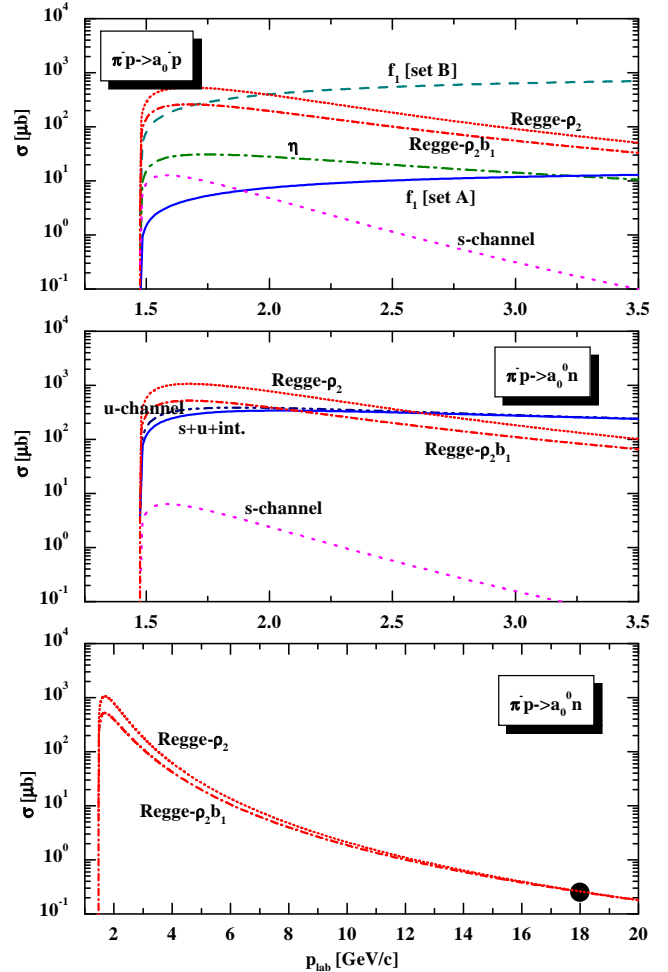,width=8.5cm}}
\caption{ The total cross sections for the reactions
$\pi^-p\rightarrow a_0^-p$ (upper part) and $\pi^-p\rightarrow
a_0^0n$ (middle and lower part) as a function of the incident
momentum. The assignment of the lines is the same as in Fig.
\protect\ref{dsdt_pip}. The experimental data point at 18 GeV/c
(lower part) is taken from Ref. \protect\cite{Brookhaven}.}
\label{stot_pip}
\end{figure}

In Fig. \ref{dsdt_pip} (upper part) we show the differential cross
sections $d\sigma/dt$ for the reaction $\pi^-p\to a_0^- p$ at 2.4
GeV/c corresponding to $\eta$ (dash-dotted) and $f_1$ exchanges
with set $A$ (solid line) and set $B$ (dashed line). A soft
cut-off parameter (set $A$) close to the mass of the $f_1$ implies
that all the contributions related to $f_1$ exchange become
negligibly small. On the other hand, for the parameter values
given by set $B$, the $f_1$ exchange contribution is much larger
than that from $\eta$ exchange. Note, that this large uncertainty
in the cut-off presently cannot be controlled by data and we will
discuss the relevance of the $f_1$ exchange contribution for all
reactions separately throughout this study. For set $B$ the total
cross section for the reaction $\pi^-p\rightarrow a_0^-p$ can be
about 0.5 mb at 2.4 GeV/c (cf. Fig.~\ref{stot_pip} (upper part))
while the forward differential cross section can be about 1
mb/GeV$^2$.


The $\eta$ and $f_1$ exchange, however, do not contribute to the
amplitude of the charge exchange reaction $\pi^-p\rightarrow
a_0^0n$. In this case we have to consider the contributions of the
$s$- and $u$-channel diagrams (Fig. \ref{Fig1} c and d):
\begin{eqnarray}
&&\hspace*{-4mm}M_N^s(\pi^-p\to a_0^0n) = g_{a_0NN} {f_{\pi NN}\over m_\pi}
\ {1\over s-m_N^2} F_N(s)
\nonumber \\
&\times&p_{1\mu}\ \bar u(p_2^\prime)  \left[(p_1+p_2)_\alpha \gamma_\alpha +m_N\right]
\gamma_\mu \ \gamma_5 u(p_2);
\label{eqpip1}
\end{eqnarray}
\begin{eqnarray}
&&\hspace*{-4mm}M_N^u(\pi^-p\to a_0^0n) = g_{a_0NN} {f_{\pi NN}\over m_\pi} \
{1\over u-m_N^2} F_N(u)  \nonumber\\
&\times&p_{1\mu} \ \bar u(p_2^\prime) \gamma_\mu \gamma_5  \left[(p_2-p_1^\prime)_\alpha
\gamma_\alpha + m_N\right] u(p_2),
\label{eqpip2}
\end{eqnarray}
where $s=(p_1+p_2)^2, \ u=(p_2-p_1^\prime)^2$ and  $m_N$ is the
nucleon mass.

The $\pi NN$ coupling constant is taken as $f_{\pi NN}^2/4\pi
=0.08$~\cite{Holinde} and the form factor for each virtual nucleon
is taken in the form \cite{Feuster}
\begin{eqnarray}
F_N(u) = \frac{\Lambda_N^4}{\Lambda_N^4+(u-m_N^2)^2}
\label{FN}\end{eqnarray}
with a cut-off parameter $\Lambda_N =1.2\div 1.3$~GeV.

The dotted and dash-double-dotted lines in the lower part of
Fig.~\ref{dsdt_pip} show the differential cross section for the
charge exchange reaction $\pi^-p\rightarrow a_0^0n$ at 2.4 GeV/c
corresponding to $s$- and $u$- channel diagrams, respectively. Due
to isospin only the $s$- channel contributes to the $\pi^-p\to
a_0^-p$ reaction (dotted line in the upper part of
Fig.~\ref{dsdt_pip}). In these calculations the cut-off parameter
$\Lambda_N$ = 1.24 GeV and $g_{a_0NN}^2/4\pi$=1.075 is taken from
Ref.~\cite{Kirchbach}. The solid line in the lower part of
Fig.~\ref{dsdt_pip} describes the coherent sum of the $s$- and
$u$- channel contributions. Except for the very forward region the
$s$- channel contribution (dotted line) is rather small compared
to the $u$- channel for the charge exchange reaction
$\pi^-p\rightarrow a_0^0n$, which may give a backward differential
cross section of about 1 mb/GeV$^2$ . The corresponding total
cross section can be about 0.3 mb at this energy (cf.
Fig.~\ref{stot_pip}, middle part).

Unfortunately, there are no experimental data for the total cross
section of $a_0$ production in $\pi N$ collisions near the threshold.
Some crude estimates can only be done by comparing the $a_0$ production
with $\rho $ and $\omega$ production. For example, the WA57
collaboration has measured inclusive photoproduction of $a_0^\pm(980)$
mesons at photon energies of 25 -- 55 GeV \cite{WA57}. It was found
that the cross section of this process is rather large and about $\sim$
1/6 of the cross sections for the corresponding non-diffractive
production of leading $\rho^0, \omega, \rho^+$ and $\rho^-$ mesons.
Furthermore, in the LBL experiment \cite{Abolins} the measured cross
sections $d\sigma /d\Omega$ for the reaction $pp\rightarrow
da_0^+(980)$ at 3.8 -- 6.3 GeV/c are $\sim (1/4\div 1/6)$ of the cross
section for $\rho ^{+}$ production (Table \ref{Tab2}).

\begin{table}[h]
\begin{center}
\begin{tabular}{ c c c c }
\hline
$pp\to d\rho^+$ & 3.8 GeV/c & 4.5 GeV/c & 6.3 GeV/c
\\
$d\sigma /d\Omega ,\mu $b$/$sr & 3.2$\pm 0.5$ & 2.0$\pm 0.4$ & 0.5$\pm 0.5$
\\[1mm]  \hline \vspace*{1mm}
$pp\to da_0^+(980)$ & 3.8 GeV/c & 4.5 GeV/c & 6.3 GeV/c
\\
$d\sigma /d\Omega ,\mu $b$/$sr & 0.5$_{-0.15}^{+0.7}$
& 0.48$_{-0.15}^{+0.28} $ & 0.35$_{-0.15}^{+0.10}$
\\ \hline
\end{tabular}
\end{center}
\caption{\label{Tab2}Cross sections for the reactions $pp\rightarrow
da_0^+(980)$ and $pp\rightarrow d\rho^+$ from
Ref.~\protect\cite{Abolins}. }
\end{table}

In view of these arguments we also compare the cross sections for
the reactions $\pi^-p\rightarrow a_0^0n$ and $\pi^-p\rightarrow
\rho^0 (\omega) n$ at 2.4 GeV/c. According to the parametrization
of Ref.~\cite{Sibirtsev} we have $\sigma(\pi^-p\rightarrow \rho^0
n) \approx 2\sigma(\pi^-p\rightarrow \omega n) \approx 1.8 \
\rm{mb}$; our estimate then gives $\sigma(\pi^-p\rightarrow
a_0^0n)\approx 0.15\div 0.3$ mb, which is in a reasonable
agreement with the $u$- channel mechanism as well as $f_1$
exchange contribution with parameters from set $B$ (cf.
Fig.~\ref{stot_pip}).

There is a single experimental point for the forward differential
cross section of the reaction $\pi^-p\rightarrow a_0^0n$ at 2.4
GeV/c (Ref. \cite{Cheshire}, lower part of Fig. 2),
$$\left.{d\sigma\over dt}(\pi^-p\rightarrow a_0^0n)\right|_{t\approx 0} =
0.49 \ \rm{mb/GeV}^2.$$ Since in the forward region ($t \approx$
0) the $s$- and $u$- channel diagrams only give a smaller cross
section, the charge exchange reaction $\pi^-p\rightarrow a_0^0n$
is most probably dominated at small $t$ by the isovector $b_1
(1^{+-})$- and $\rho_2 (2^{--})$- meson exchanges (see e.g. Ref.
\cite{Achasov}). Though the couplings of these mesons to $\pi a_0$
and $NN$ are not known, we can estimate
$\frac{d\sigma}{dt}(\pi^-p\rightarrow a_0^0n)$ in the forward
region using the Regge-pole model as developed by Achasov and
Shestakov \cite{Achasov}. Note, that the Regge-pole model is
expected to provide a reasonable estimate for the cross section at
medium energies of about a few GeV and higher (see e.g. Refs.
\cite{Kaidalov1,Kondrat} and references therein).

\subsection{The Regge-pole model}

The $s$- channel helicity amplitudes for the reaction $\pi^-p
\rightarrow a_0^0n$ can be written as
\begin{eqnarray}
&&\hspace*{-4mm}M_{\lambda_2^\prime\lambda_2}(\pi^-p\rightarrow a_0^0n) =
\bar u_{\lambda_2^\prime}(p_2^\prime)\ \left[-A(s,t) \right.\nonumber\\
&+&\left.(p_1+p_1^\prime)_\alpha \gamma_\alpha {B(s,t)\over 2}\right]
\gamma_5 u_{\lambda_2}(p_2),
\label{Reg1}\end{eqnarray}
where the invariant amplitudes $A(s,t)$ and $B(s,t)$ do not contain
kinematical singularities and (at fixed $t$ and large $s$) are related
to the helicity amplitudes as
\begin{eqnarray}
M_{++}\approx -sB, \hspace{3mm} M_{+-}\approx \ M_{++}\approx
\sqrt{t_{\min }-t}\ A. \label{Reg2}\end{eqnarray}
The differential
cross section then can be expressed through the helicity
amplitudes in the standard way as
\begin{eqnarray}
\hspace*{-4mm} {d\sigma\over dt}(\pi^-p\rightarrow a_0^0n)
={1\over 64 \pi s} {1\over (p_1^{\rm{cm}})^2}  (|M_{++}|^2+|M_{+-}|^2).
\label{eq:sigQGSM1}
\end{eqnarray}

Usually it is assumed that the reaction $\pi^-p\rightarrow a_0^0n$
at high energies is dominated by the $b_1$ Regge-pole exchange.
However, as shown by Achasov and Shestakov \cite{Achasov} this
assumption is not compatible with the angular dependence of
$d\sigma/dt(\pi^-p \rightarrow a_0^0n)$ observed at Serpukhov at
40 GeV/c \cite{Serpukhov,Serpukhov1} and Brookhaven at 18 GeV/c
\cite{Brookhaven}. The reason is that the $b_1$ Regge trajectory
contributes only to the amplitude $A(s,t)$ giving a dip in
differential cross section at forward angles, while the data show
a clear forward peak in $d\sigma/dt(\pi^-p\rightarrow a_0^0n)$ at
both energies. To interpret this phenomenon Achasov and Shestakov
introduced a $\rho_2$ Regge-pole exchange conspiring with its
daughter trajectory.  Since the $\rho_2$ Regge trajectory
contributes to both invariant amplitudes, $A(s,t)$ and $B(s,t)$,
its contribution does not vanish at $\Theta =0$ thus giving a
forward peak due to the term $|M_{++}|^2$ in $d\sigma/dt$. At the
same time the contribution of the $\rho _2$ daughter trajectory to
the amplitude $A(s,t)$ is necessary to cancel the kinematical pole
at $t=0$ introduced by the $\rho_2$ main trajectory (conspiracy
effect). In this model the $s$- channel helicity amplitudes can be
expressed through the $b_1$ and the conspiring $\rho_2$ Regge
trajectories exchange as
\begin{eqnarray}
M_{++}\approx M_{++}^{\rho_2}(s,t) = \gamma_{\rho_2}(t)
\exp [-i {\pi\over 2} \alpha_{\rho _2}(t)]
\left( \frac s{s_0}\right)^{\alpha_{\rho_2}(t)}\hspace*{-6mm},
\hspace*{6mm} \label{Reg3}\end{eqnarray}
\begin{eqnarray}
M_{+-}\approx M_{+-}^{b_1}(s,t) &=& \sqrt{(t_{\min }-t)/s_0}\
\gamma_{b_1}(t) \nonumber\\
&\times& i\exp [-i {\pi\over 2} \alpha_{b_1}(t)]
\left(\frac s{s_0}\right)^{\alpha_{b_1}(t)},
\label{Reg4}\end{eqnarray}
where $\gamma _{\rho_2}(t) = \gamma_{\rho _2}(0)\ \exp (b_{\rho_2}t)$,
$\gamma_{b_1}(t)\ = \gamma_{b_1}(0)\ \exp (b_{b_1}t)$,
$t_{\min}\approx -m_N^2 (m_{a_0}^2-m_\pi ^2)/s^2$,
$s_0\approx 1$ GeV$^2$ while the meson Regge trajectories have the
linear form $\alpha_j(t) = \alpha_j(0)+\alpha_j^\prime(0)t$.

Achasov and Shestakov describe the Brookhaven data on the $t$
distribution at 18~GeV/c for $-t_{\min }\leq -t\leq 0.6$ GeV$^2$
\cite{Brookhaven} by the expression
\begin{eqnarray}
\frac{dN}{dt} = C_1 \left[e^{\Lambda_1t} + (t_{\min }-t)
\frac{C_2}{C_1} e^{\Lambda_2t} \right], \label{Reg5}\end{eqnarray}
where the first and second terms describe the $\rho _2$ and $b_1$
exchanges, respectively. They found two fits: a) $\Lambda_1=4.7$
GeV$^{-2}, C_2/C_1=0$; b) $\Lambda_1=7.6$ GeV$^{-2},
C_2/C_1\approx 2.6$~GeV$^{-2}, \Lambda _2=5.8$~GeV$^{-2}$.
This implies that the $b_1$ contribution is equal to zero for fit a)
and yields only 1/3 of the integrated cross section for fit b) at 18
GeV/c.  Moreover, using the available data on the reaction
$\pi^-p\rightarrow a_2^0(1320)n$ at 18 GeV/c and comparing them with
the data on the $\pi^-p\rightarrow a_0^0n$ reaction they estimated the
total and forward differential cross sections $\sigma
(\pi^-p\rightarrow a_0^0n\rightarrow \pi ^0\eta n)\approx 200$ nb and
$[d\sigma /dt(\pi^-p\rightarrow a_0^0n\rightarrow \pi^0\eta
n)]_{t=0}\approx 940$ nb/GeV$^2.$ Taking $Br(a_0^0\rightarrow
\pi^0\eta)\approx 0.8$ we find $\sigma (\pi^-p\rightarrow
a_0^0n)\approx 0.25$ $\mu$b and $[d\sigma /dt(\pi^-p\rightarrow
a_0^0n)]_{t=0}\approx 1.2$ $\mu$b/GeV$^2$.  In this way all the
parameters of the Regge model can be fixed and we will employ it for
the energy dependence of the $\pi^-p\rightarrow a_0^0n$ cross section
to obtain an estimate at lower energies, too.

The mass of the $\rho_2(2^{--})$ is expected to be about 1.7 GeV
(see \cite{Kokoski} and references therein) and the slope of the
meson Regge trajectory in the case of light ($u, d$) quarks is 0.9
GeV$^{-2}$ \cite{Kaidalov}. Therefore, the intercept of the
$\rho_2$ Regge trajectory is $\alpha_{\rho
_2}(0)=2-0.9m_{\rho_2}^2\approx -0.6$. Similarly -- in the case of
the $b_1$ trajectory -- we have $\alpha_{b_1}(0)\approx -0.37$. At
forward angles we can neglect the contribution of the $b_1$
exchange (see discussion above) and write the energy dependence of
the differential cross section in the form
\begin{eqnarray}
\left.{d\sigma_{Regge}\over dt}(\pi^{-}p\rightarrow a_0^0n)\right|_{t=0}
&\approx&\left.\frac{d\sigma_{\rho_2}}{dt}\right|_{t=0} \sim\nonumber\\
&\sim&{1\over (p_1^{cm})^2}\left(\frac s{s_0}\right)^{-2.2}\hspace*{-5mm}.
\hspace*{5mm} \label{eq:sigQGSM2}
\end{eqnarray}
This provides the following estimate for the forward differential
cross section at 2.4 GeV/c,
\begin{eqnarray}
\left.{d\sigma_{Regge}\over dt} (\pi^-p\rightarrow
a_0^0n)\right|_{t=0} \approx 0.6\ \rm{ mb/GeV}^2,
\label{eq:sigQGSM3}
\end{eqnarray}
which is in agreement with the experimental data point
\cite{Cheshire} (lower part of Fig. 2). Since the $b_1$ and
$\rho_2$ Regge trajectories have isospin 1, their contribution to
the cross section for the reaction $\pi^-p\rightarrow a_0^-p$ is
twice smaller,
\begin{eqnarray}
{d\sigma_{Regge}\over dt} (\pi^-p\rightarrow a_0^-p)=
\frac 12\ {d\sigma_{Regge}\over dt} (\pi^-p\rightarrow a_0^0n).
\label{eq:sigQGSM4}
\end{eqnarray}

In Fig. \ref{dsdt_pip} the short-dotted lines  show the resulting
differential cross sections for $d\sigma_{Regge}(\pi^-p\rightarrow
a_0^-p)/dt$ (upper part) and $d\sigma_{Regge}(\pi^-p\rightarrow
a_0^0n)/dt$ (lower part) at 2.4 GeV/c corresponding to $\rho_2$
Regge exchange (fit a) ), whereas the dash-dotted lines indicate the
contribution for $\rho_2$ and $b_1$ Regge trajectories (fit b) ).
For $t\to 0$  both Regge parametrizations agree, however, at large
$|t|$ the solution including the $b_1$ exchange gives a smaller cross
section. The cross section $d\sigma_{Regge}(\pi^-p\rightarrow
a_0^-p)/dt$ in the forward region exceeds the contributions of $\eta$,
$f_1$ (set $A$) and $s$- channel exchanges, however, is a few times
smaller than the $f_1$ exchange contribution for set $B$. On the other
hand, the cross section $d\sigma_{Regge} (\pi^-p\rightarrow a_0^0n)/dt$
is much larger than the $s$- and $u$- channel contributions in the
forward region, but much smaller than the $u$- channel contribution in
the backward region.

The integrated cross sections for $\pi^- p \rightarrow a_0^- p$
(upper part) and $\pi^- p \rightarrow a_0^0 n$ (middle and lower
part) for the Regge model are shown in Fig. \ref{stot_pip} as a
function of the pion lab. momentum by short-dotted lines for
$\rho_2$ exchange and by short dash-dotted lines for $\rho_2, b_1$
trajectories. In the few GeV region the cross sections are
comparable with the $u$-channel and $f_1$ -exchange contribution
(set $B$). At higher energies it decreases as $s^{-3.2}$ in
contrast to the non-Reggeized $u$-channel and $f_1$- exchange
contributions which anyhow should only be employed close to the
threshold region.

The main conclusions of this Section are as follows: In the region
of a few GeV the dominant mechanisms of  $a_0$ production in the
reaction $\pi N \rightarrow a_0 N$ are  $u$-channel nucleon and
$t$-channel $f_1$ -meson exchanges which give cross sections for
$a_0$ production about $0.3 \div 0.4$ mb (cf. upper part of
Fig.~\ref{stot_pip}).  Similar cross sections ($\simeq 0.4 \div 1$
mb) are predicted by the Regge model with conspiring $\rho_2$ (or
$\rho_2$ and $b_1$) exchanges, normalized to the Brookhaven data
at 18 GeV/c (lower part of Fig.~\ref{stot_pip}). The contributions
of $s$-channel nucleon and $t$-channel $\eta$ -meson exchanges are
small (cf. upper and middle parts of Fig.~\ref{stot_pip}).

\section{The reaction $pp \to d a_0^+$}
\label{sec:da0}

The missing mass spectrum in the reaction $pp\rightarrow dX$
for deuterons produced at $0^{\circ}$ in the laboratory and
incident momenta of 3.8, 4.5 and 6.3 GeV/c has been measured at
LBL (Berkeley) \cite {Abolins}. It is interesting, that apart from
the missing mass peaks corresponding to $\pi$ and $\rho$
production, there is a distinctive structure in the missing mass
spectrum at 0.95 GeV$^2$, which was identified as $a_0$
production.

\begin{figure}[t]
\centerline{\psfig{file=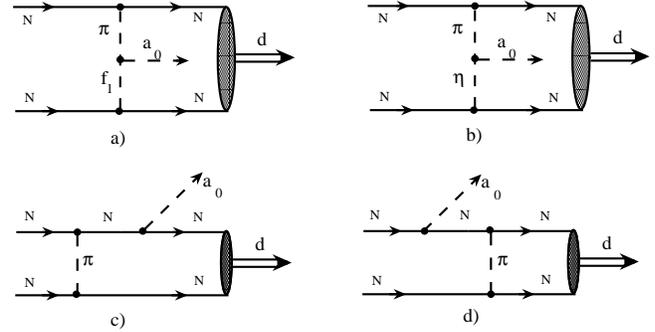,width=8.5cm}}
\caption{ The diagrams describing the different mechanisms of the
$a_0$-meson production in the reaction $NN\to da_0$ within the
framework of the two-step model (TSM).}
\label{fig:tsm}
\end{figure}

In order to estimate the cross section for the reaction $pp \to d
a_0^+$ at lower momenta (available at COSY) we use the two-step
model (TSM) (cf. Refs.~\cite{Grishina1,Grishina2}). The
contributions of hadronic intermediate states to the $P$-wave
amplitude of the reaction $pp\to da_0^{+}$ (within the framework
of the TSM) are described by the diagrams $a - d$ in
Fig.~\ref{fig:tsm}. We consider three different contributions from
the amplitude $\pi N \to a_0 N$:  i) the $f_1(1285)$- meson
exchanges (Fig.~\ref{fig:tsm} a); ii) the $\eta$- meson exchange
(Fig.~\ref{fig:tsm} b); iii) $s$- and $u$- channel nucleon
exchanges (Fig.~\ref{fig:tsm} c and d). As follows  from the
analysis in Sect. 3 the contributions of the $\eta$- exchange and
$s$- channel nucleon can be neglected. We thus restrict to the
$f_1$- exchange (set $B$) and  the $u$- channel nucleon current.

The cut-off $\Lambda_N$ for nucleon exchange  (Eq.~(\ref{FN})) is
considered as a free parameter now within the interval 1.2 -- 1.3
GeV. In order to preserve the correct structure of the amplitude
under permutations of the initial nucleons (which are
antisymmetric in the isovector state) the amplitude is written as
the difference of $t$- and $u$- channel contributions in the form
\begin{equation}
T_{pp\to dM}^\pi (s,t,u) = A_{pp\to dM}(s,t)-A_{pp\to dM}(s,u),
\label{Atu}
\end{equation}
where $M$ stands for the $a_0^+$- meson. Furthermore,
$s=(p_1+p_2)^2$, $t=(p_3-p_1)^2$, $u=(p_3-p_2)^2$ where $p_1$,
$p_2$, $p_3$, and $p_4$ are the 4-momenta of the initial protons,
meson $M$ and the deuteron, respectively. The structure of the
amplitude (\ref{Atu}) guarantees that the S-wave part vanishes
since it is forbidden by angular momentum conservation and the
Pauli principle.

Since we are interested in the $pp\to da_0^{+}$ cross section near
threshold, where the momentum of the deuteron is comparatively
small, we use a non-relativistic description of this particle by
neglecting the 4th component of it's polarization vector.
Correspondingly, the relative motion of nucleons in the deuteron
is also treated non-relativistically. Then one can write the first
($t$-channel) term on the r.h.s. of Eq.~(\ref{Atu}) as
(\cite{Grishina1})
\begin{eqnarray}
&&\hspace*{-4mm} A_{pp\to da_0^+}(s,t) = \frac{f_{\pi NN}}{m_\pi} \
g_{f_1NN}\ g_{f_1a_0\pi} \label{Adir} \\
&\times& \sqrt{(p_1^0+m_N)(p_2^0+m_N)}  \nonumber \\
&\times& M^{jl}({\vec p}_1,{\vec p}_3)\ \varphi_{\lambda_2}^T(\vec p_2)\
(-i\sigma_2)\sigma^j  \vec\sigma \cdot {\vec\epsilon}^{\ *(d)} \sigma^l
\varphi_{\lambda_1}(\vec p_1), \nonumber
\end{eqnarray}
where $\vec \epsilon^{\ (d)}$ is the polarization vector of the
deuteron; $p_1^0=p_2^0=\sqrt{\vec p_1^{\,2}+m_N^2}$, while
$\varphi_\lambda $ are the (2-component) spinors of the nucleons
in the initial state. The tensor function ${M}^{jl}({\vec
p}_1,{\vec p}_3)$ is defined by the integral
\begin{eqnarray}
&&\hspace*{-4mm} M^{jl}(\vec p_1,\vec p_3) = \sqrt{2 m_N} \int \
\frac{d^3p_2^\prime}{(2\pi)^{3/2}} \label{Mr}\\
&\times& \sqrt{(p_1^{\prime 0}+m_N)
(p_2^{\prime 0}+m_N)}\ \left\{\frac{p_1^j}{p_1^0+m_N} +
\frac{p_2^{\prime j}}{p_2^{\prime 0}+m_N}\right\} \nonumber \\
&\times& I \cdot \Phi_{\pi^0N\to a_0^0N}^l
(\vec p_2^{\, \prime},\vec p_1,\vec p_3)\
\frac{F_{\pi NN}(q_\pi^2)}{q_\pi ^2-m_\pi ^2}\
\Psi_d(\vec p_2^{\, \prime} + \vec {p_3}/2), \nonumber
\end{eqnarray}
where the contribution of $f_1$- exchange is given by
\begin{eqnarray}
&&\hspace*{-4mm}\Phi_{\pi^0N\to a_0^0N(f_1)}^l (\vec p_2^{\, \prime},\vec p_1,\vec p_3) =
g_{f_1NN}\ g_{f_1a_0\pi}\ \frac{F_{f_1NN}(q_{f_1}^2)} {q_{f_1}^2-m_{f_1}^2}
\nonumber\\
&\times& \left\{ 2p_3^l-\frac{2(p_3+p_2^\prime)^l}{p_1^{\prime 0}+m_N}\
\left(m_N\left[1+\frac{m_{a_0}^2-q_2^2}{m_{f_1}^2}\right] -p_3^0\right)
\right.\nonumber\\
&-&\left.\frac{2p_1^l}{p_1^0+m_N}\ \left(m_N\left[1
+\frac{m_{a_0}^2-q_2^2}{m_{f_1}^2}\right] + p_3^0\right) \right\}.
\label{eqPhi1}
\end{eqnarray}
The $u$- channel contribution reads
\begin{eqnarray}
&&\hspace*{-4mm}\Phi_{\pi^0N\to a_0^0N(u)}^l
(\vec p_2^{\, \prime},\vec p_1,\vec p_3)
=g_{a_0NN}\ \frac{f_{\pi NN}}{m_\pi}\ 2m_N
 \nonumber \\
&\times&\left\{-p_3^l+\frac{(p_3+p_2^\prime)^l}{p_1^{\prime \,0}+m_N}\
\left( {m_N\over 2}\ \left[3+\frac{q_N^2}{m_N^2}\right] -p_3^0\right)\right.
 \label{eqPhi2}\\
&+&\left.\frac{p_1^l}{p_1^0+m_N}\ \left( {m_N\over2}\
\left[3+\frac{q_N^2}{m_N^2}\right] +p_3^0\right)\right\}
\ \frac{F_N(q_N^2)}{q_N^2-m_N^2}.
 \nonumber
\end{eqnarray}
Here $\Psi_d(\vec p_2^{\,\prime} +\vec{p_3}/2)$ is the deuteron wave
function for which we use the Paris model \cite{Lacombe}.
In (\ref{Mr}) $I$ is the isospin factor
which depends on the mechanism of the reaction $pp\to (pn)a_0^+$.
For $f_1$ and $u$- channel exchange we have
$I_{(f_1)}=1$ and $I_{(u)}=3\sqrt{2}$, respectively.
Further kinematical quantities, which also dependent on the
momenta $\vec p_1$, $\vec p_3$ and $\vec p_2^\prime$, are defined
as
\begin{eqnarray}
&& q_\pi^2 = -\delta_0(\vec p_2^{\, \prime 2}+\beta_\pi(\vec p_1))
 - 2\vec p_1 \vec p_2^{\, \prime}, \nonumber\\
&& q_{f_1}^2 = -\delta_0\left(\vec p_2^{\, \prime 2}+\beta_{f_1}
(\vec p_1,\vec p_3)\right) + \frac{p_3^0}{m_N}\ \vec p_2^{\, \prime 2}
  \nonumber\\
&&\hspace*{6mm} -2\vec p_1\cdot \vec p_2^{\, \prime}-2\vec p_3\cdot
\vec p_2^{\, \prime} -2\vec p_3\cdot \vec p_1, \nonumber \\[1mm]
&& q_N^2 = m_N^2+m_{a_0}^2-2p_1^0p_3^0+2\vec p_1\cdot \vec p_3,  \nonumber\\
&& \beta_\pi(\vec p_1) = (\vec p_1^{\, 2}-T_1^2)/\delta _0, \label{eqkin}\\
&& \beta_{f_1}(\vec p_1,\vec p_3) = \beta_\pi(\vec p_1)-m_{a_0}^2/\delta_0
+p_3^0m_N \nonumber \\
&& \delta_0 = p_1^0/m_N,\  \ T_1=\sqrt{\vec p_1^{\, 2}+m_N^2}-m_N\ , \nonumber\\
&& p_2^{ \prime \, 0} = \sqrt{\vec p_2^{\, \prime 2}+m_N^2}, \ \
p_3^0 = \sqrt{\vec p_3^{\, 2}+m_{a_0}^2},  \nonumber \\
&& p_1^{ \prime 0} = \sqrt{(\vec p_2^{\, \prime} +\vec p_3)^{\, 2}+m_N^2}.
 \nonumber
\end{eqnarray}
with $m_{a_0}$ denoting the mass of the $a_0$ meson. The form
factors $F_{f_1 NN}$ and $F_{\pi NN}$ are taken in the form
(\ref{form}) within $\Lambda_{\pi NN} = 1.3$~GeV for the $\pi NN$
vertex according to  Ref. \cite{Holinde} and parameter set $B$ for
the $f_1 NN$ vertex. The $u$- channel term $A_{pp\to da_0^+}(s,u)$
in Eq.~(\ref{Atu}) can be obtained from (\ref{Adir}) by the
substitution $p_1\leftrightarrow p_2$,
$\varphi_{\lambda_1}\leftrightarrow \varphi_{\lambda_2}$.

\begin{figure}[h]
\centerline{\psfig{file=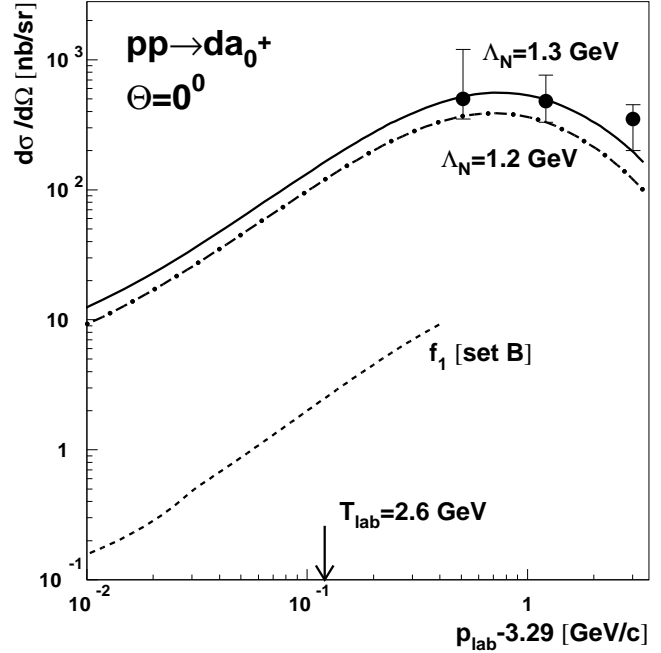,width=8.5cm}}
\caption{Forward differential cross section of the reaction
$pp\rightarrow d a_0^+$ as a function of $(p_{\rm{lab}} - 3.29)$
GeV/c. The full dots are the experimental data from
Ref.~\protect\cite{Abolins}. The dash-dotted and solid lines
describe the results of the TSM calculated at $\Lambda_N = 1.2$
and 1.3 GeV, respectively. The dotted line shows the
contribution of $f_1$ exchange for the parameter set $B$ (see
text).} \label{fig:dsacosy}
\end{figure}

The differential cross section $pp\to da_0^+$ then can be written
as
\begin{eqnarray}
{d\sigma_{pp\to da_0^+}\over dt} &=&
{1\over 64\pi s}\ \frac 1{(p_1^{\rm{cm}})^2} \label{eq:sigmaVd}\\
&\times&\overline{|A_{pp\to da_0^{+}}(s,t) - A_{pp\to da_0^{+}}(s,u)|^2}.
\nonumber
\end{eqnarray}
The calculated forward differential cross section  (as a function
of the proton-beam momentum) is presented in
Fig.~\ref{fig:dsacosy}. The dash-dotted and solid lines describe
the results of the TSM for different values of the nucleon cut-off
parameter: $\Lambda _N=1.2$ and 1.3 GeV, respectively. A rather
good description of the existing data \cite{Abolins} is achieved
for $\Lambda _N=1.3$ GeV (solid line). We recall that in Sect. 3
we have used $\Lambda_N =$ 1.24 GeV from Ref. \cite{Kirchbach}
which gives a cross section in between the dash-dotted and solid
line. Our predictions for this cross section practically do not
depend on the couplings of the $f_1$ since the $f_1$ exchange
contribution turns out to be very small even for parameter set $B$
(dashed line in Fig.~\ref{fig:dsacosy}).  The arrow in
Fig.~\ref{fig:dsacosy} indicates the maximum proton momentum
presently available at COSY. At this energy a differential cross
section of $0.1 \div 0.2$ $\mu$b/sr should be expected according to
our calculations.

\begin{figure}[t]
\centerline{\psfig{file=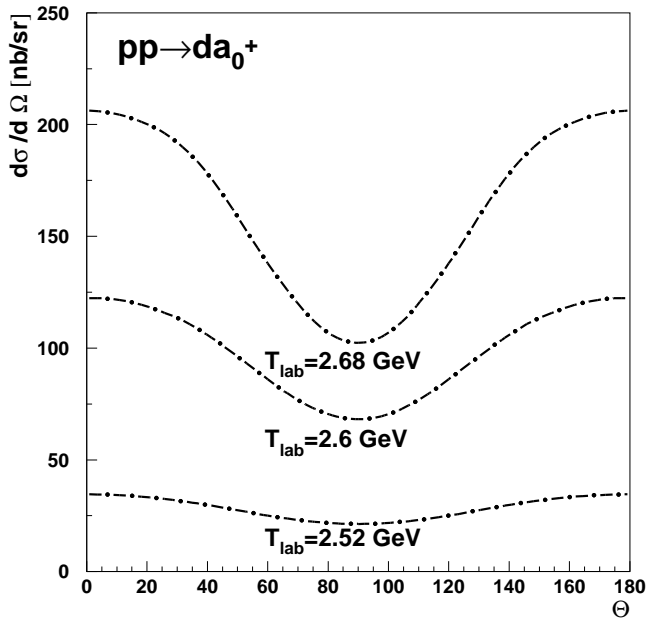,width=8.5cm}}
\caption{Angular dependence of the differential cross section
$d\sigma/d\Omega$ of the reaction $pp\rightarrow d a_0^+$ in
the c.m.s. at different energies. The cut-off parameter for the
$u$-channel nucleon exchange is $\Lambda_N = 1.3$ GeV.}
\label{fig:dsacosy1}
\end{figure}

In Fig.~\ref{fig:dsacosy1} the calculated angular differential
cross section for the reaction $pp\to da_0^{+}$ is shown as a
function of the center-of-mass angle $\Theta$ which can be
parametrized as
\begin{eqnarray}
\frac{d\sigma }{d\Omega} = A+B\cdot \cos^2{\Theta}
+ C\cdot \cos^4{\Theta}.
\label{eqTheta}
\end{eqnarray}
The results of our calculations in the framework of the TSM for
$\Lambda_N=1.3$ GeV  are:\\
$A=21.3$ nb/sr, $B=15.3$ nb/sr, $C=-2.1$ nb/sr \\
 at $T_{\rm{lab}}=2.52$ GeV ($\sigma_{tot}=330$ nb);\\
$A=68$ nb/sr, $B=76$ nb/sr, $C=-22$ nb/sr \\
 at $T_{\rm{lab}}=2.6$ GeV  ($\sigma_{tot}=1120$ nb);\\
$A=78$ nb/sr, $B=97$ nb/sr, $C=-31$ nb/sr \\
 at $T_{\rm{lab}}=2.62$ GeV ($\sigma_{tot}=1310$ nb).

We note that an understanding of the $a_0(980)$ production
mechanism may also give interesting information on its internal
structure. For example, the WA57 collaboration has interpreted the
relatively strong production of the $a_0^\pm(980)$ in photon
induced reactions at energies of 25 -- 55 GeV as evidence for a
$q\bar q$ state rather than a $qq\bar q\bar q$ state \cite{WA57}.
This argument can also be used here. If measurements at COSY will
confirm a comparatively large value of the $a_0^+(980)$- production
cross section as presented in this work, this will provide further
evidence that the $a_0^+(980)$ has an essential admixture of a
$q\bar q$ component.

\section{Conclusions}
\label{sec:Conclusion}

In this work  we have estimated $a_0$ production cross sections in
the reaction $\pi N \to a_0 N$ near threshold and at medium
energies by considering the $a_0(980)$-resonance  as a usual $q
\bar q$-meson. Using an effective Lagragian approach we have
analyzed different contributions to the differential and total
cross sections, i.e. $t$- channel $\eta$- and $f_1$- meson
exchanges as well as $s$- and $u$-channel nucleon exchanges, and
have found that the $f_1$- and $u$- channel contributions are
dominant in the $\pi^- p \to a_0^- p$ and $\pi^- p
\rightarrow a_0^0 n$ reactions, respectively. We have analyzed
also predictions of the Regge model with conspiring $\rho_2$
exchange normalized to the data at 18 GeV/c. We found that this
model gives (in the few GeV region) a cross section comparable to
the $f_1$- and $u$- channel mechanisms.

The latter results have been used to calculate the differential
and total cross section of the reaction $p p \to d a_0^+$ within
the framework of the two-step model, where the amplitude of the
$NN \to d a_0$ reaction can be expressed through the amplitude of
the $\pi N \to a_0 N$ reaction and a structure integral containing
the deuteron wave function in the non-relativistic limit. It is
found that the cross section of the $p p \to d a_0^+$ reaction is
dominated almost entirely by the $u$- channel mechanism reaching
a value of about 1 $\mu$b at $T_{lab}$ = 2.6 GeV. An experimental
confirmation of this comparatively large production cross section
would imply that the $a_0^+(980)$ has an essential admixture of a
$q\bar q$ component.

\section*{Acknowledgements}
We are grateful to  A. Sibirtsev for helpful discussions.


\end{document}